\documentclass[journal]{IEEEtran}
\usepackage{cite}

\ifCLASSINFOpdf
\usepackage[pdftex]{graphicx}
\DeclareGraphicsExtensions{.pdf,.jpeg,.png}
\else
\usepackage[dvips]{graphicx}
\DeclareGraphicsExtensions{.eps}
\fi
\usepackage{epsfig,epsf,epstopdf,graphicx}
\usepackage[cmex10]{amsmath}
\usepackage{amssymb}
\usepackage{amsthm }
\usepackage{nicefrac}
\usepackage{hyperref}
\usepackage{xcolor}
\hypersetup{colorlinks=true}
\usepackage{tabularx}

\theoremstyle{theorem}

\theoremstyle{definition}

\theoremstyle{plain}

\theoremstyle{plain}

\newcommand\Tstrut{\rule{0pt}{2.6ex}}         
\newcommand{\vb}{{\textbf{v}}}

\usepackage{algorithm}
\usepackage{multirow}
\usepackage{algcompatible}
\usepackage[bottom]{footmisc}

\PassOptionsToPackage{dvipsnames}{xcolor}
\usepackage{tikz}
\usetikzlibrary{calc}

\ifCLASSOPTIONcompsoc
  \usepackage[caption=false,font=normalsize,labelfont=sf,textfont=sf]{subfig}
\else
  \usepackage[caption=false,font=footnotesize]{subfig}
\fi
\usepackage{fixltx2e}
\usepackage{afterpage}
\usepackage{float}
\usepackage{stfloats}

\makeatletter

\begin{document}

%
\title{Resistor Hopping KLJN Noise Communication Using Small Bias Voltages Supported by ML and Optimum Threshold-Based Detectors}

\author{Hadi~Zayyani, \IEEEmembership{Member,~IEEE,} Felipe A. P. de Figueiredo, Mohammad Salman, Rausley A. A. de Souza, \IEEEmembership{Senior Member,~IEEE}


\thanks{H.~Zayyani is with the Department
of Electrical and Computer Engineering, Qom University of Technology (QUT), Qom, Iran (e-mails: zayyani@qut.ac.ir, motie.s@qut.ac.ir).}
\thanks{M.~Salman is with College of Engineering and Technology, American University of the Middle East, Egaila, 54200, Kuwait (e-mail: mohammad.salman@aum.edu.kw).}
\thanks{F. A. P. de Figueiredo, and R. A. A. de Souza are with the National Institute of Telecommunications (Inatel), Santa Rita do Sapucaí, Brazil. (e-mail: felipe.figueiredo@inatel.br, rausley@inatel.br).}



\vspace{-0.5cm}}


\maketitle
\thispagestyle{plain}
\pagestyle{plain}

\begin{abstract}
In this paper, a Resistor Hopping (RH) scheme with the addition of biases is proposed for secure Kirchhoff-Law Johnson-Noise (KLJN) communication. The RH approach enables us to increase the bit rate of secure communication between Alice and Bob, while also ensuring that the inherent unconditional security of KLJN is satisfied. The biases are added to the proposed scheme to better distinguish between Gaussian-distributed noises in terms of their means, rather than just using variances. Throughout the paper, we strive to minimize biases to achieve a power-efficient scheme. For the detection part of the proposed algorithm, a Maximum-Likelihood (ML) detector is derived. The separability condition of Gaussian distributions is investigated, along with the provision of a threshold-based detector that offers both simple and optimal thresholds in terms of minimizing the error probability. Some analysis of the proposed RH-KLJN communication scheme is provided, including Physical Layer Security (PLS) equations. Simulation results demonstrate the advantages of the proposed scheme over the classical KLJN scheme, offering a higher data rate and lower bit error probability at the expense of increased complexity.
\end{abstract}


%
\IEEEpeerreviewmaketitle

\section{Introduction}
\label{sec:Intro}

\IEEEPARstart{N}{oise} communication is a promising secure communication system with zero power \cite{Basar23}. Its roots date back to the early 2000s, with the concept of modulated thermal noise \cite{Basar23}. The noise communication scheme addresses two serious needs for the next generation of communication systems, such as 6G and beyond \cite{Basar21}. The first is the need for low-power-consuming schemes for communication systems or radically zero-power communication systems. This requisite may decrease the complexity and even the expenditure of a communication system as well. The noise communication approach radically enables the use of zero-power communication devices. The second is the need for a secure communication system. Noise communication inherently provides unconditional security \cite{Basar23}.

The concept of noise communication is well studied in the works by Kish \cite{Kish05}-\cite{Kish17book}. In pioneering work \cite{Kish05}, Kish proposed the concept of a zero-power stealthy communication system from an applied physics perspective, utilizing two resistors with different impedances. In a sequel, the same author proposed a Kirchhoff-Law-Johnson Noise (KLJN) secure key exchange scheme, in which the laws of physics provide the basis for unconditionally secure communication: Kirchhoff's law and thermal noise of two pairs of resistors \cite{Kish06}. Moreover, Kish et al. proposed totally secure classical networks with multipoint telecloning (teleportation) of classical bits through loops with Johnson-like noise \cite{Kish06_1}. The KLJN communication scheme utilizes two different levels (high and low) of resistors to represent bits 1 and 0 of Alice and Bob. Then, Alice and Bob share a common wire in which the common voltage represents a zero-mean Gaussian random variable with three different variance levels based on four different bit cases of Alice and Bob. Both partners of Alice and Bob easily confirm the case of $00$ and $11$. But, this situation is not secure since Eve can detect the information bits of both ends. Therefore, Alice and Bob discard this case. Instead, when the cases of $01$ or $10$ occur, the variance of the common voltage has a seminal value, which, even if it is detected by Eve, she can not retrieve the information bits of Alice and Bob. In contrast, each agent can detect the partner's bit by flipping its own bit. Then, it is said that unconditional security is achieved, which is equivalent to the quantum secrecy level. In addition, Kish provided a very powerful and simple encryption method, in which the location of 0 and 1 bits cannot be determined from measurements taken between terminals, known as the Kish cipher \cite{Kish17book}. On the other side of research, the Index Modulation (IM) concept \cite{Basar17}-\cite{Basar19} is suggested, where the idea is similar to KLJN noise communication, in which indexing is utilized to embed information into entities, just as with the indexing of two resistors \cite{Basar23}.

More advanced works on KLJN noise communication have been presented in the past 15 years. In \cite{Ming08}, the feasibility of this scheme for communication in the ranges of two to two thousand kilometers is investigated. Moreover, \cite{Kish10} discusses the effect of wire resistance on the noise voltage and current. In \cite{Saez13}, the first attempt to calculate the Bit Error Probability (BEP) of the KLJN noise communication scheme is performed by considering only the common voltage. Afterwards, for further reduction of BEP, both voltage and current noises are utilized \cite{Saez13_1}. Then, \cite{Smul14} used a more advanced method for calculating the BEP of the KLJN communication system. In addition, noise properties in a KLJN communication system are investigated in \cite{Ging14}. In a rather interesting work, a generalized KLJN approach is proposed in which Alice and Bob use different low and high resistors \cite{Vadai15}. So, in this generalized scheme, four resistors are utilized. Later, \cite{Ferd20} demonstrates the lower security of the generalized KLJN scheme than the classical KLJN scheme in realistic situations. Moreover, two works are presented on security and performance analysis of KLJN and generalized KLJN secure key exchanger protocol \cite{Ming15}, \cite{Ming17}. Besides, \cite{Vadai16} proposed a generalized attack protection in the KLJN Secure Key Exchanger. In addition, an experimental realization of an ultralow-power wireless communication method is presented that works by selectively connecting or disconnecting an impedance-matched resistor and an antenna \cite{Kape22}. Additionally, from a communication engineering perspective, Basar recently formulated a new framework for the BEP calculation of the KLJN noise communication system and suggested two novel detectors that further reduce the BEP \cite{Basar23}. Recently, \cite{Tasci25} proposed a Flip-KLJN secure noise communication where a pre-agreed intermediate level, such as high/low (H/L), triggers a flip of the bit map value during the bit exchange period. More recently, a KLJN-based thermal noise modulation has been proposed, which represents a viable solution for secure Internet of Things (IoT) communication at ultra-low power levels \cite{Salem25}. It proposes a new asymmetric KLJN-based modulation scheme that utilizes a four-resistor structure to improve the bit error rate (BER) without requiring additional noise samples per bit. In a recent paper \cite{Basar24}, Basar proposed a noise modulation scheme in which the noise samples with different variances are directly fed to the wireless antenna. In that work, the Bit Error Probability (BEP) is calculated in closed form, and the Noise Modulation system performance is evaluated under fading channels. Also, optimal detection and performance analysis of a thermal noise modulation is investigated in \cite{Alshaw24}. Recently, an on-off digital modulation noise is presented in \cite{Anjos25} and an innovative joint energy harvesting and communication scheme is suggested for future Internet-of-Things (IoT) devices by leveraging the emerging noise modulation technique \cite{Yapici25}.

As can be seen from the communication engineering perspective of Basar in \cite{Basar23}, it appears that noise communication is again under the focus of the communication community. Although it has been around for around 20 years in the literature, it is still in its infancy in the communication community. Many tools of communication theory can be generalized to the noise communication framework. In this paper, inspired by Frequency Hopping (FH) communication theory, we propose using a Resistor Hopping (RH) communication system. This new scheme increases the secure data rate between Alice and Bob. It inherently reduces the BEP by using two small bias voltages in the low- and high-resistance parts of the KLJN framework which is similar to what presented in \cite{Yapici25} but for different purpose of better distinguishability of Gaussian distributions rather than energy harvesting. The idea of resistor hopping is to divide the bit duration into several chip durations, and in each chip duration, sub-bits of Alice and Bob can be considered. In each chip duration, based on their sub-bit, Alice and Bob switch (hop) between two low resistors (sub-bit 0 or 1) in the presence of the main bit 0 and between two high resistors (sub-bit 0 or 1) in the presence of the main bit 1. Although this RH scheme employs four resistors, as in the generalized KLJN scheme \cite{Vadai15}, it exhibits superior performance compared to KLJN, as confirmed by simulations. The simulation results demonstrate the superior performance of RH-KLJN in terms of lower BEP at the same sampling rate. It also supports a much secure data rate than KLJN.

The paper is organized as follows. After the literature survey in the introduction, Section~\ref{sec:ProblemForm} discusses the system model and problem formulation. In Section~\ref{sec: prop}, the proposed RH framework is presented and developed. Section \ref{sec: det} drives some basic detectors for the proposed scheme, and the overall algorithm is developed. Some analysis, especially from the viewpoint of Physical Layer Security (PLS), is presented in section \ref{sec: ana}. Simulation results are presented in Section~\ref{sec: Simulation}, while conclusions are drawn in Section~\ref{sec: con}.

\section{System Model and Preliminaries}
\label{sec:ProblemForm}
In this section, we present the basics of the secure KLJN noise communication scheme. This scheme is based on the Johnson noise voltages generated by two terminals, Alice and Bob, which are connected through a wire channel. Alice and Bob do not use any modulated signal; instead, they utilize the thermal noise of their connecting resistors. When Alice's bit or Bob's bit is 0, they connect the common wire to low resistors. Otherwise, if their bit is 1, they connect the common wire to high resistors. Thus, the resistor of Alice is $R_A\in\{R_L,R_H\}$ when her bit is 0 or 1. Similarly, the resistor of Bob is $R_B\in\{R_L,R_H\}$ when his bit is 0 or 1. Therefore, based on the two scenarios presented by Alice and Bob, four cases arise. In the $00$ case, the common voltage has a Gaussian distribution of zero-mean and variance equal to \cite{Basar23}:
\begin{align}
\sigma^2_{00}=4kT \frac{R_L}{2}\Delta f,
\end{align}
where $k=1.38\times 10^{-23}$ is the Boltzmann's constant, $T$ is the temperature of resistors, and $\Delta f$ is the bandwidth \cite{Basar23}. In the $11$ case, the common voltage is zero-mean Gaussian with variance equal to \cite{Basar23}:
\begin{align}
\sigma^2_{11}=4kT \alpha\frac{R_L}{2}\Delta f,
\end{align}
where $\alpha=\frac{R_H}{R_L}$. In the cases of $01$ and $10$, the common voltage is Gaussian distributed with zero-mean and variance equal to \cite{Basar23}:
\begin{align}
\sigma^2_{01}=\sigma^2_{10}=4kT \frac{\alpha}{\alpha+1}R_L\Delta f.
\end{align}
The KLJN detector (on either side of Alice or Bob) estimates the variance, and since we have $\sigma^2_{11}>\sigma^2_{01}>\sigma^2_{00}$, by distinguishing from the different variances, he (she) obtains the state. The states of $00$ and $11$ are not secure since Eve can detect the bits of both Alice and Bob. In these two cases, Alice and Bob discard the bits. However, in the cases of $01$ or $10$, there is an unconditionally quantum level security in which Eve can not distinguish between the two cases. Still, fortunately, any agent can detect the partner's bit by flipping their bit to exchange a bit. The disadvantage of this KLJN scheme is that distinguishing the variance value is somewhat challenging and prone to error, as the variance estimator is more prone to error than the mean estimator. In the proposed RH scheme, presented in the next section, we utilize bias voltages to distinguish between the Gaussian distributions and their respective means.

\section{Proposed Resistor Hopping Noise Communication Scheme}
\label{sec: prop}
In this section, we propose the Resistor Hopping KLJN (RH-KLJN) noise communication scheme. The basic idea is that each duration bit $T_b$ of Alice and Bob is divided into $P$ sub-bits (See Figure 1). In the sub-bit duration, $T_c=\frac{T_b}{P}$, the high or low resistor is randomly switched between two high resistors $R_{H_0}$ and $R_{H_1}$, assuming $R_{H_1}=\beta R_{H_0}>R_{H_0}$, or two low resistors $R_{L_0}$ and $R_{L_1}$, assuming $R_{L_1}=\beta R_{L_0}>R_{L_0}$, based on random sub-bits. As for the classic paper of KLJN, we assume $R_{H_i}=\alpha R_{L_i}$ for $i=0,1$. We also assume that $\alpha>\beta$. The block diagram and system model are shown in Figures 2 and 3. The high resistors are selected as $R_{H_i}\in\{R_{H_0},R_{H_1}\}$ and $R_{L_i}\in\{R_{L_0},R_{L_1}\}$. Moreover, the actual resistors used in the scheme are selected as $R_A\in\{R_{L_0}, R_{L_1}, R_{H_0}, R_{H_1}\}$ and $R_B\in\{R_{L_0}, R_{L_1}, R_{H_0}, R_{H_1}\}$ based on the main bit and sub-bit of Alice and Bob. In figure 2, we also add two biases, $m_L$ and $m_H=\gamma m_L>m_L$, to the thermal noises of the resistors in the low- and high-resistor branches. This will result in Gaussian noise with non-zero mean. As we will see later, this results in the scheme's capability to distinguish between Gaussian distributions and detect bits within the sub-bit duration. It is a small energy-consuming scheme and enabling low-power noise communication instead of zero-power noise communication. As confirmed by simulation results, the performance of bit detection in the presence of these biases is better than that of the classical KLJN scheme with zero mean. Also, the proposed RH with biases increases the data rate by adding sub-bits in the bit duration.

\begin{figure}[h]
\begin{center}
\includegraphics[scale=0.16]{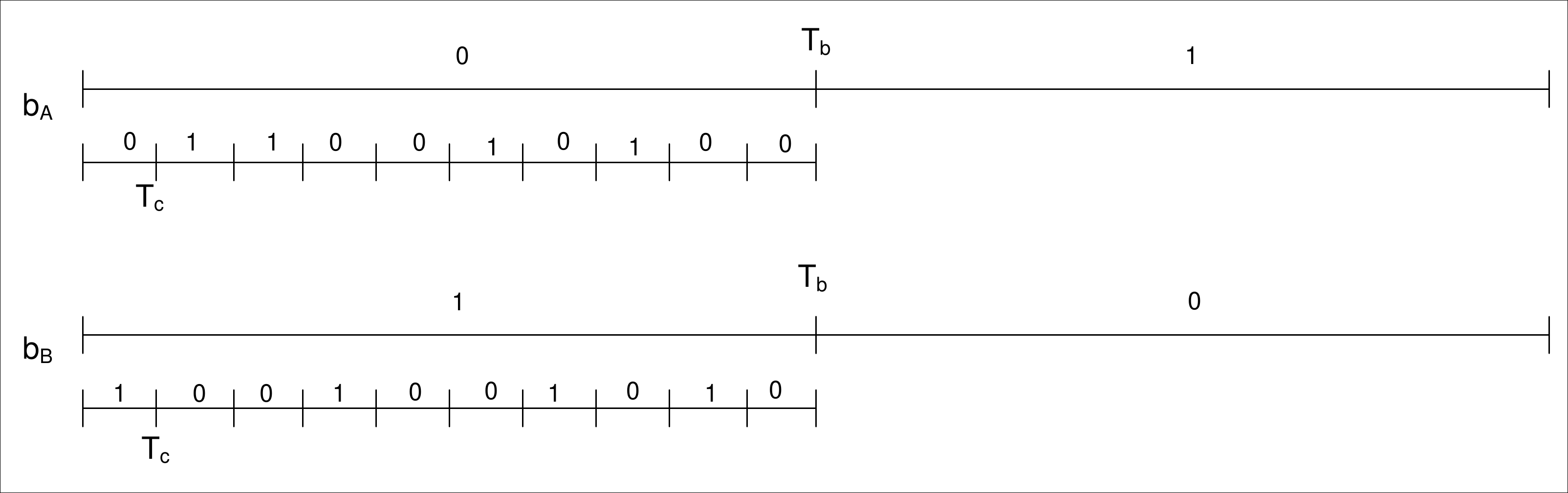}
\end{center}
\vspace{-0.5 cm}
\caption{Bit duration and chip duration of resistor hopping scheme.}
\label{fig1}
\end{figure}

\begin{figure}[h]
\begin{center}
\includegraphics[scale=0.4]{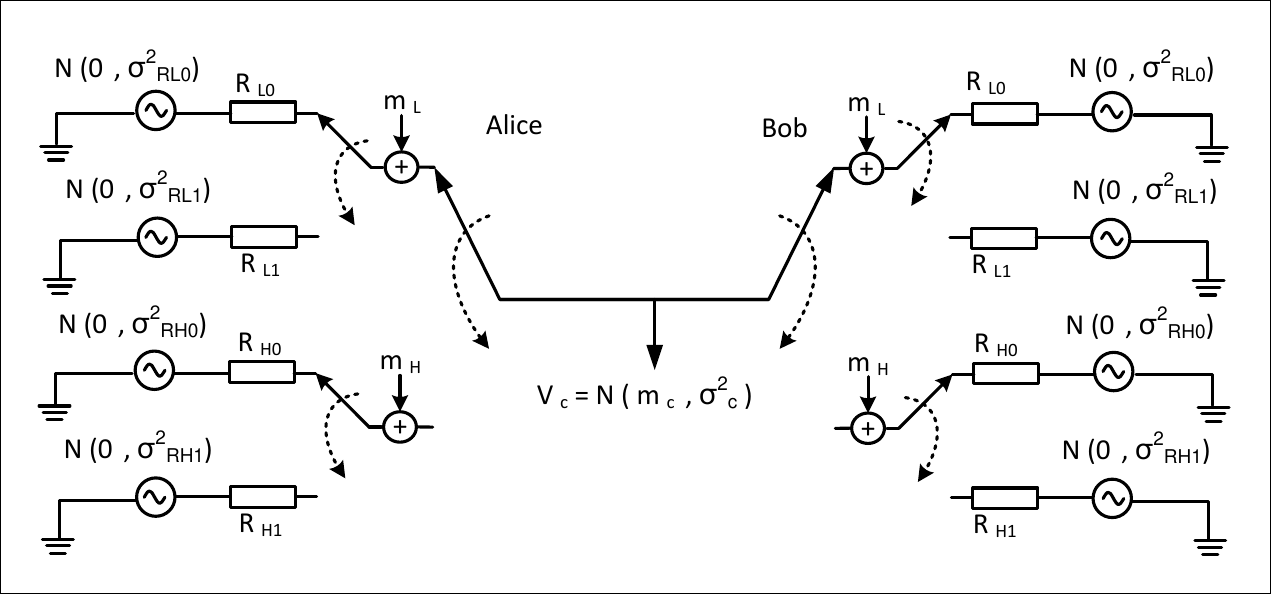}
\end{center}
\vspace{-0.5 cm}
\caption{Block diagram of resistor hopping KLJN scheme.}
\label{fig2}
\end{figure}

\begin{figure}[h]
\begin{center}
\includegraphics[scale=0.43]{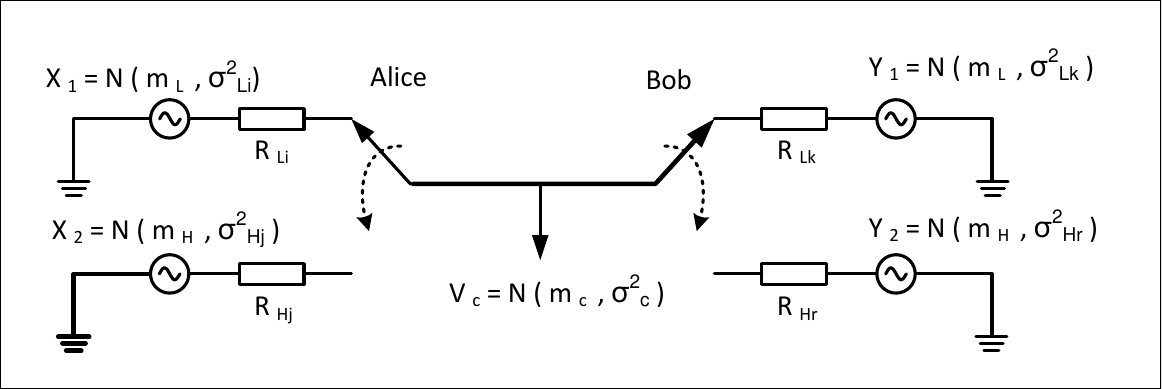}
\end{center}
\vspace{-0.5 cm}
\caption{System model of resistor hopping KLJN scheme.}
\label{fig3}
\end{figure}

In the sequel, we analyze the different cases of bits of Alice and Bob. We nominate the current bit of Alice as $b_A$ and the bit of Bob as $b_B$. We consider four different cases for Alice and Bob bits, which are as follows.

\subsection{Case of $(b_A,b_B)=(0,0)$}
In this case, the scheme uses the low resistors in both Alice and Bob. In each sub-bit duration, Alice randomly (with uniform probability of 0.5) uses $R_{L_i}$ for $i=0,1$ and Bob uses similarly randomly (but independent of Alice) $R_{l_k}$ for $k=0,1$. Then, the common voltage at the wire connecting Alice and Bob is a random Gaussian variable $v_c$ with mean $m_c$ and variance $\sigma^2_c$, which are given by:
\begin{align}
v_c=x_1\frac{R_{L_k}}{R_{L_k}+R_{L_i}}+y_1\frac{R_{L_i}}{R_{L_i}+R_{L_k}},
\end{align}
\begin{align}
m_c=m_L\frac{R_{L_k}}{R_{L_k}+R_{L_i}}+m_L\frac{R_{L_i}}{R_{L_i}+R_{L_k}}=m_L,
\end{align}
\begin{align}
\sigma^2_c=\Big(\frac{R_{L_k}}{R_{L_k}+R_{L_i}}\Big)^2\sigma^2_{L_i}+\Big(\frac{R_{L_i}}{R_{L_i}+R_{L_k}}\Big)^2\sigma^2_{L_k},
\end{align}
where $x_1$ and $y_1$ are shown in Figure~1, $\sigma^2_{L_i}=4KTR_{L_i}\Delta f$, $\sigma^2_{L_k}=4KTR_{L_k}\Delta f$, and the resistors $R_{L_i}$, and $R_{L_k}$ are switched between $R_{L_0}$ and $R_{L_1}$ when Alice's sub-bit and Bob's sub-bit are $b_{A(or B),p}=0$ and $b_{A(or B),p}=1$, in which $1\le p\le P$ is the index of sub-bit. Based on each sub-bit of Alice $b_{A,p}=0,1$ and Bob $b_{B,p}=0,1$ in the $p$'th chip of each bit duration, we have four cases which mean and variance of common Gaussian voltage are computed with some manipulations and are shown in Table~I, in which we defined $a\triangleq 4KT\Delta f$. In this case, we have four Gaussian cases, two of which are identical. In this case, the common voltage has a mixture of three Gaussian distributions with probabilities $0.25$, $0.5$, and $0.25$, respectively, which all have the same mean $m_L$ with different variances $\frac{1}{2}aR_{L_0}$, $aR_{L_0}||R_{L_1}$, and $\frac{1}{2}aR_{L_1}$, respectively, where $R_1||R_2\triangleq\frac{R_1R_2}{R_1+R_2}$ is the parallel resistor of two resistors $R_1$ and $R_2$.

\begin{table}[!b]
\caption{Mean and variance of Gaussian random variable of common voltage $v_c$ in $p$'th chip time interval and in the case of $b_A=0$ and $b_B=0$}
\centering
 \begin{tabular}{p{14mm}||p{25mm}|p{30mm}} \hline
 \Tstrut $(i,k)=(b_{A,p},b_{B,p})$ & Mean & Variance \\\hline \hline

 \Tstrut (0,0)
 & $\!\!\! \begin{aligned} m_L& \end{aligned} $	
 &	$\!\!\!\begin{aligned} \frac{1}{2}aR_{L_0}& \end{aligned}$  \\ \hline

 \Tstrut (0,1)
 &	$\!\!\!\begin{aligned} m_L& \end{aligned}$	
 & $\!\!\!\begin{aligned} aR_{L_0}||R_{L_1}=\frac{a\beta}{\beta+1}R_{L_0}& \end{aligned}$ \\ \hline

 \Tstrut (1,0)
 &	$m_L$	
 &  $aR_{L_0}||R_{L_1}=\frac{a\beta}{\beta+1}R_{L_0}$ \\ \hline

 (1,1)
 &	$m_L$	
 &	$\frac{1}{2}aR_{L_1}$ \\ \hline

\end{tabular}
 \begin{tabular} {l}
\\
\end{tabular}
\label{Table_1}
\end{table}

\subsection{Case of $(b_A,b_B)=(0,1)$}
In this case, the scheme uses the low resistor for Alice and the high resistor for Bob. In each sub-bit duration, Alice randomly (with uniform probability of 0.5) uses $R_{L_i}$ for $i=0,1$ and Bob uses similarly randomly (but independent of Alice) $R_{H_r}$ for $r=0,1$. This way, the common voltage at the wire connecting Alice and Bob is a random Gaussian variable $v_c$ with mean $m_c$ and variance $\sigma^2_c$, which are given by:
\begin{align}
v_c=x_1\frac{R_{H_r}}{R_{H_R}+R_{L_i}}+y_2\frac{R_{L_i}}{R_{L_i}+R_{H_r}},
\end{align}
\begin{align}
m_c=m_L\frac{R_{H_r}}{R_{H_r}+R_{L_i}}+m_H\frac{R_{L_i}}{R_{L_i}+R_{H_r}},
\end{align}
\begin{align}
\sigma^2_c=\Big(\frac{R_{H_r}}{R_{H_r}+R_{L_i}}\Big)^2\sigma^2_{L_i}+\Big(\frac{R_{L_i}}{R_{L_i}+R_{H_r}}\Big)^2\sigma^2_{H_r},
\end{align}
where $x_1$ and $y_2$ are shown in Figure~1, $\sigma^2_{L_i}=aR_{L_i}$, $\sigma^2_{H_r}=aR_{H_r}$, and the resistors $R_{L_i}$, are switched between $R_{L_0}$ and $R_{L_1}$ when Alice's are $b_{A,p}=0$ or $b_{A,p}=1$, respectively, and $R_{H_r}$ are switched between $R_{H_0}$ and $R_{H_1}$ when Bob's sub-bit are $b_{B,p}=0$ and $b_{B,p}=1$. Based on each sub-bit of Alice $b_{A,p}=0,1$ and Bob $b_{B,p}=0,1$ in the $p$'th chip of each bit duration, we have four cases, for which the mean and variance of common Gaussian voltage are computed with some manipulations and are shown in Table~II. In this case, we have three Gaussian distributions with different means and different variances. Therefore, the common voltage consists of a mixture of three Gaussian distributions, where two of them have probabilities of $0.25$ each, and the other has a probability of $0.5$. Simple inspection of inequalities show that $c_1>c_2$ for all $\gamma>1$, and hence $m_1>m_2$. Additionally, by inspection of the inequalities, we have $c_2<c_3$ for all $\beta>1$, which implies $m_3>m_2$. Moreover, we see that $c_3>c_1$ for all $\gamma>1$ and hence $m_3>m_1$. Hence, we have $m_2<m_1=m_4<m_3$.

\begin{table}[!b]
\caption{Mean and variance of Gaussian random variable of common voltage $v_c$ in $p$'th chip time interval and in the case of $b_A=0$ and $b_B=1$}
\centering
 \begin{tabular}{p{14mm}||p{37mm}|p{22mm}} \hline
 \Tstrut $(i,r)=(b_{A,p},b_{B,p})$ & Mean & Variance \\\hline \hline

 \Tstrut (0,0)
 & $\!\!\! \begin{aligned} m_1=&m_L\frac{R_{H_0}}{R_{H_0}+R_{L_0}}+\\&m_H\frac{R_{L_0}}{R_{L_0}+R_{H_0}}=c_1m_L\end{aligned} $	
 &	$\!\!\!\begin{aligned} \sigma^2_1&=aR_{L_0}||R_{H_0}\\ &=\frac{a\alpha R_{L_0}}{\alpha+1}& \end{aligned}$  \\ \hline

 \Tstrut (0,1)
 &	$\!\!\!\begin{aligned} m_2=&m_L\frac{R_{H_1}}{R_{H_1}+R_{L_0}}+\\&m_H\frac{R_{L_0}}{R_{L_0}+R_{H_1}}=c_2m_L& \end{aligned}$	
 & $\!\!\!\begin{aligned} \sigma^2_2&=aR_{L_0}||R_{H_1}\\ &\frac{a\alpha\beta R_{L_0}}{\alpha\beta+1} \end{aligned}$ \\ \hline

 \Tstrut (1,0)
 &	$\!\!\!\begin{aligned} m_3=&m_L\frac{R_{H_0}}{R_{H_0}+R_{L_1}}+\\&m_H\frac{R_{L_1}}{R_{L_1}+R_{H_0}}=c_3m_L& \end{aligned}$	
 & $\!\!\!\begin{aligned} \sigma^2_3&=aR_{L_1}||R_{H_0}\\ &\frac{a\alpha\beta R_{L_0}}{\alpha+\beta} \end{aligned}$ \\ \hline

 (1,1)
 &	$\!\!\!\begin{aligned} m_4=&m_L\frac{R_{H_1}}{R_{H_1}+R_{L_1}}+\\&m_H\frac{R_{L_1}}{R_{L_1}+R_{H_1}}=c_4m_L& \end{aligned}$		
 & $\!\!\!\begin{aligned} \sigma^2_4&=\sigma^2_1=aR_{L_1}||R_{H_1}\\ &\frac{a\alpha R_{L_0}}{\alpha+1} \end{aligned}$ \\ \hline

\end{tabular}
 \begin{tabular} {l}
\\
$c_1\triangleq\frac{\alpha+\gamma}{\alpha+1}$, $c_2\triangleq\frac{\gamma+\alpha\beta}{\alpha\beta+1}$, $c_3\triangleq\frac{\alpha+\gamma\beta}{\alpha+\beta}$, $c_4=c_1=\triangleq\frac{\alpha+\gamma}{\alpha+1}$
\end{tabular}
\label{Tablef_2}
\end{table}

\subsection{Case of $(b_A,b_B)=(1,0)$}
In this case, the scheme uses a high resistor for Alice and a low resistor for Bob. In each sub-bit duration, Alice randomly (with uniform probability of 0.5) uses $R_{H_j}$ for $j=0,1$ and Bob uses similarly randomly (but independent of Alice) $R_{L_k}$ for $k=0,1$. So, the common voltage at the wire connecting Alice and Bob is a random Gaussian variable $v_c$ with mean $m_c$ and variance $\sigma^2_c$, which are given by:
\begin{align}
v_c=x_2\frac{R_{L_k}}{R_{L_k}+R_{H_j}}+y_1\frac{R_{H_j}}{R_{H_j}+R_{L_k}},
\end{align}
\begin{align}
m_c=m_H\frac{R_{L_k}}{R_{L_k}+R_{H_j}}+m_L\frac{R_{H_j}}{R_{H_j}+R_{L_k}},
\end{align}
\begin{align}
\sigma^2_c=\Big(\frac{R_{L_k}}{R_{L_k}+R_{H_j}}\Big)^2\sigma^2_{H_j}+\Big(\frac{R_{H_j}}{R_{H_j}+R_{L_k}}\Big)^2\sigma^2_{L_k},
\end{align}
where $x_2$ and $y_1$ are shown in Figure~1, $\sigma^2_{L_k}=aR_{L_k}$, $\sigma^2_{H_j}=aR_{H_j}$, and the resistors $R_{L_k}$, are switched between $R_{L_0}$ and $R_{L_1}$ when Bob's sub-bit is $b_{B,p}=0$ or $b_{B,p}=1$, respectively, and $R_{H_j}$ are switched between $R_{H_0}$ and $R_{H_1}$ when Alice's sub-bit are $b_{A,p}=0$ and $b_{A,p}=1$. Based on each sub-bit of Alice $b_{A,p}=0,1$ and Bob $b_{B,p}=0,1$ in the $p$'th chip of each bit duration, we have four cases, for which the mean and variance of common Gaussian voltage are computed with some manipulations and are shown in Table~III. In this case, we have again three Gaussian distributions with different means and different variances. In this case of $(b_A,b_B)=(1,0)$, the means and variances are the same as the previous case of $(b_A,b_B)=(0,1)$. We will discuss the Gaussian distributions in more detail in the future, once all cases have been investigated.

\begin{table}[!b]
\caption{Mean and variance of Gaussian random variable of common voltage $v_c$ in $p$'th chip time interval and in the case of $b_A=1$ and $b_B=0$}
\centering
 \begin{tabular}{p{14mm}||p{37mm}|p{22mm}} \hline
 \Tstrut $(j,k)=(b_{A,p},b_{B,p})$ & Mean & Variance \\\hline \hline

 \Tstrut (0,0)
 & $\!\!\! \begin{aligned} m_1=&m_H\frac{R_{L_0}}{R_{L_0}+R_{H_0}}+\\&m_L\frac{R_{H_0}}{R_{H_0}+R_{L_0}}=c_1m_L& \end{aligned} $	
 &	$\!\!\!\begin{aligned} \sigma^2_1=aR_{L_0}||R_{H_0}& \end{aligned}$  \\ \hline

 \Tstrut (0,1)
 &	$\!\!\!\begin{aligned} m_3=&m_H\frac{R_{L_1}}{R_{L_1}+R_{H_0}}+\\&m_L\frac{R_{H_0}}{R_{H_0}+R_{L_1}}=c_3m_L& \end{aligned}$	
 & $\!\!\!\begin{aligned} \sigma^2_3=aR_{H_0}||R_{L_1}& \end{aligned}$ \\ \hline

 \Tstrut (1,0)
 &	$\!\!\!\begin{aligned} m_2=&m_H\frac{R_{L_0}}{R_{L_0}+R_{H_1}}+\\&m_L\frac{R_{H_1}}{R_{H_1}+R_{L_0}}=c_2m_L& \end{aligned}$	
 &  $\sigma^2_2=aR_{L_0}||R_{H_1}$ \\ \hline

 (1,1)
 &	$\!\!\!\begin{aligned} m_4=&m_H\frac{R_{L_1}}{R_{L_1}+R_{H_1}}+\\&m_L\frac{R_{H_1}}{R_{H_1}+R_{L_1}}=c_1m_L& \end{aligned}$		
 &	$\sigma^2_1=aR_{L_1}||R_{H_1}$ \\ \hline

\end{tabular}
 \begin{tabular} {l}
\\
\end{tabular}
\label{Table_3}
\end{table}

\subsection{Case of $(b_A,b_B)=(1,1)$}
In this case, the scheme uses the high resistors in both Alice and Bob. In each sub-bit duration, Alice randomly (with uniform probability of 0.5) uses $R_{H_j}$ for $j=0,1$ and Bob uses similarly randomly (but independent of Alice) $R_{H_r}$ for $r=0,1$. So, the common voltage at the wire connecting Alice and Bob is a random Gaussian variable $v_c$ with mean $m_c$ and variance $\sigma^2_c$, which are:
\begin{align}
v_c=x_2\frac{R_{H_r}}{R_{H_r}+R_{H_j}}+y_2\frac{R_{H_j}}{R_{H_j}+R_{H_r}},
\end{align}
\begin{align}
m_c=m_H\frac{R_{H_r}}{R_{H_r}+R_{H_j}}+m_H\frac{R_{H_j}}{R_{H_j}+R_{H_r}}=m_H,
\end{align}
\begin{align}
\sigma^2_c=\Big(\frac{R_{H_r}}{R_{H_r}+R_{H_j}}\Big)^2\sigma^2_{H_j}+\Big(\frac{R_{H_j}}{R_{H_j}+R_{H_r}}\Big)^2\sigma^2_{H_r},
\end{align}
where $x_2$ and $y_2$ are shown in Figure~1, $\sigma^2_{H_j}=aR_{H_j}$, $\sigma^2_{H_r}=aR_{H_r}$, and the resistors $R_{H_j}$, and $R_{H_r}$ are switched between $R_{H_0}$ and $R_{H_1}$ when Alice's sub-bit and Bob's sub-bit are $b_{A(or B),p}=0$ and $b_{A(or B),p}=1$, where $1\le p\le P$ is the index of sub-bit. Based on each sub-bit of Alice $b_{A,p}=0,1$ and Bob $b_{B,p}=0,1$ in the $p$'th chip of each bit duration, we have four cases, for which the mean and variance of the common Gaussian voltage are computed with some manipulations and are shown in Table~IV. In this case, we have four Gaussian cases, two of which are identical. In this case, the common voltage has a mixture of three Gaussian distribution with probabilities $0.25$, $0.5$, and $0.25$, respectively, which all have the same mean of $m_H$ with different variances of $\frac{1}{2}aR_{H_0}$, $aR_{H_0}||R_{H_1}$, and $\frac{1}{2}aR_{H_1}$, respectively.

\subsection{General Discussion}
As mentioned before, in each sub-bit duration, there are 16 different cases for the main bits of $(b_A,b_B)$ and for the sub-bits of $(b_{A,p},b_{B,p})$. The means and variances of the common voltage in these 16 cases are shown in Tables I-IV. We summarize the 16 cases in Table V, presenting the mean for each case. After inspecting these cases, we observe that there are 9 different Gaussian distributions. If we sketch the distributions, as shown in Figure~4, three Gaussian distributions are in the lower level with mean equal to $m_L$, three Gaussian distributions are in the middle level, and 3 Gaussian distributions are in the higher level with mean equal to $m_H$. Therefore, we can say that, in general, the common voltage has a mixture of 9 Gaussian distributions. However, within each sub-bit duration, there is a single Gaussian distribution with mean and variance values reported in Tables I-IV, with the mean listed in Table V.

\begin{table}[!b]
\caption{Mean and variance of Gaussian random variable of common voltage $v_c$ in $p$'th chip time interval and in the case of $b_A=1$ and $b_B=1$}
\centering
 \begin{tabular}{p{14mm}||p{30mm}|p{30mm}} \hline
 \Tstrut $(j,r)=(b_{A,p},b_{B,p})$ & Mean & Variance \\\hline \hline

 \Tstrut (0,0)
 & $\!\!\! \begin{aligned} m_H& \end{aligned} $	
 &	$\!\!\!\begin{aligned} \frac{1}{2}aR_{H_0}& \end{aligned}$  \\ \hline

 \Tstrut (0,1)
 &	$\!\!\!\begin{aligned} m_H& \end{aligned}$	
 & $\!\!\!\begin{aligned} aR_{H_0}||R_{H_1}& \end{aligned}$ \\ \hline

 \Tstrut (1,0)
 &	$m_H$	
 &  $aR_{H_0}||R_{H_1}$ \\ \hline

 (1,1)
 &	$m_H$	
 &	$\frac{1}{2}aR_{H_1}$ \\ \hline

\end{tabular}
 \begin{tabular} {l}
\\
\end{tabular}
\label{Table_4}
\end{table}

\begin{figure}[h]
\begin{center}
\includegraphics[width=9cm]{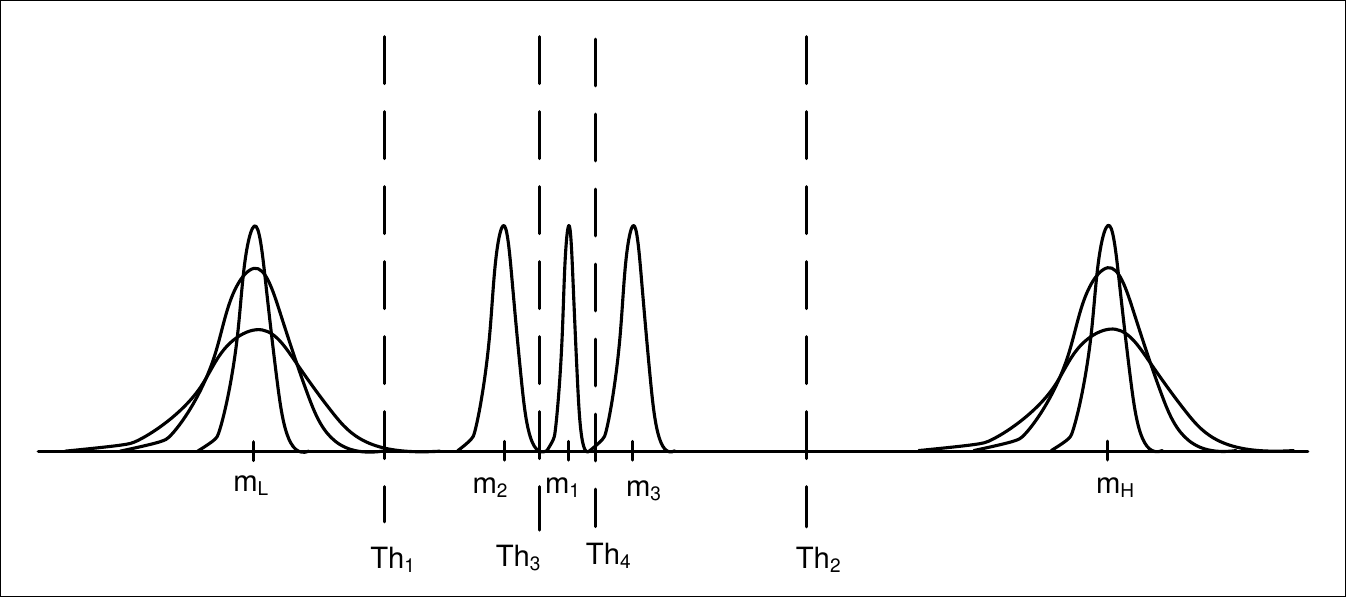}
\end{center}
\vspace{-0.5 cm}
\caption{Gaussian mixture distributions for the common voltage in the resistor hopping scheme.}
\label{fig3}
\end{figure}

\begin{table}[!b]
\caption{Mean and variance of Gaussian random variable of common voltage $v_c$ in $p$'th chip time interval. $m_1$, $m_2$, $m_3$ and $m_4$ are defined in Table II. The bold notations are related to the cases of unconditionally security.}
\centering
 \begin{tabular}{p{14mm}||p{12mm}|p{12mm}|p{12mm}|p{12mm}} \hline
 \Tstrut $(b_{A,p},b_{B,p})$ & (00) & (01)& (10) & (11) \\\hline \hline

 \Tstrut $(b_A,b_B)=(00)$
 & $\!\!\! \begin{aligned} m_L\end{aligned} $	
 & $\!\!\!\begin{aligned} m_L& \end{aligned}$	
 & $\!\!\!\begin{aligned} m_L& \end{aligned}$	
 &	$\!\!\!\begin{aligned} m_L& \end{aligned}$  \\ \hline

 \Tstrut $(b_A,b_B)=(01)$
 &	$\!\!\!\begin{aligned} m_1& \end{aligned}$	
 & $\!\!\!\begin{aligned} \mathbf{m_2}& \end{aligned}$
 & $\!\!\!\begin{aligned} \mathbf{m_3}& \end{aligned}$
 & $\!\!\!\begin{aligned} m_4 \end{aligned}$ \\ \hline

 \Tstrut $(b_A,b_B)=(10)$
 &	$\!\!\!\begin{aligned} m_1& \end{aligned}$	
 &	$\mathbf{m_3}$	
 &	$\mathbf{m_2}$	
 & $\!\!\!\begin{aligned} m_4 \end{aligned}$ \\ \hline

 $(b_A,b_B)=(11)$
 &	$\!\!\!\begin{aligned} m_H& \end{aligned}$		
 &	$m_H$	
 &	$m_H$	
 & $\!\!\!\begin{aligned} m_H \end{aligned}$ \\ \hline

\end{tabular}
 \begin{tabular} {l}
\\
\end{tabular}
\label{Table_5}
\end{table}

The main point is that the three Gaussian cases in the lower level can not be distinguished (or are at least hard to distinguish) since they have the same mean of $m_L$. The main advantage of the proposed scheme in this paper is that it utilizes biases, which enable us to separate the Gaussian distributions by their means rather than just using their variances. If two Gaussian distributions have the same mean (for example, zero in the classic scheme in KLJN), the separation is based on observing different variances, which is difficult when the variances are near each other. Also, a similar setting is satisfied for the three Gaussian distributions in the higher level with mean equal to $m_H$. However, fortunately, the three middle-level Gaussian distributions, which are indeed related to 8 of the total 16 cases, have different means that can be distinguished from each other if their means are sufficiently separated based on their variances. We will discuss the ideal separability of Gaussian distributions in the next section. Another basic point is that these 8 cases are for the cases of $(b_{A},b_{B})=(0,1)$ and $(b_{A},b_{B})=(1,0)$, which pairwise have the same Gaussian distribution. A deliberate inspection of Table V reveals that we should discard the cases of $(b_A,b_B)= (00) $ and $(b_A,b_B)= (11) $ entirely. On the other hand, for the cases $(b_A,b_B)=(01)$ and $(b_A,b_B)=(10)$, we should discard the cases $(b_{A,P},b_{B,p})=(00)$ and $(b_{A,P},b_{B,p})=(11)$. Thus, the same unconditional security is there in the sub-bit domain. Hence, we only exchange the bits in the 4 cases out of the 16 cases depicted in Table V. Fortunately, similar to the unconditional secrecy of the KLJN scheme, we have the same condition here.

\section{Gaussian detector and bit detector for the RH-KLJN scheme}
\label{sec: det}
In this section, we investigate detectors for the RH-KLJN algorithm. Two main detectors are proposed: the first is a simple minimum distance detector, and the second is the Maximum-Likelihood (ML) detector. In both detectors, we initially identify the Gaussian distributions that are far away from the three middle-level distributions. This is necessary since they are the cases that should be discarded in the bit exchange. Fortunately, this is simple since they are far apart from the three middle Gaussian distributions. We use two thresholds, $\mathrm{Th}_1$ and $\mathrm{Th}_2$, as depicted in Figure~4. To determine the values of these thresholds, and for the separability condition, we should assume that the low tail of the rightmost of the three middle Gaussians is far apart from the high tail of the three Gaussians at low levels. Then, we should have:
\begin{align}
\label{eq: cond}
m_L+3\sqrt{\frac{a\alpha}{2}R_{L_0}}\ll m_2-3\sigma_2.
\end{align}

Some simplification shows that this condition is satisfied by using a proper, large enough bias of $m_L$ as
\begin{align}
m_L\gg\frac{3\sqrt{0.5a\alpha R_{L_0}}+3\sqrt{0.5a\frac{\alpha\beta}{\alpha\beta+1}R_{L_0}}}{c_2-1}=k_1\sigma_{R_{L0}},
\end{align}
where $\sigma_{R_{L_0}}=\sqrt{aR_{L_0}}$, and the coefficient $k_1$ is equal to
\begin{align}
k_1\triangleq\frac{3(\sqrt{0.5\alpha}+\sqrt{\frac{0.5\alpha\beta}{\alpha\beta+1}})}{c_2-1}=\frac{3(\sqrt{0.5\alpha}+\sqrt{\frac{0.5\alpha\beta}{\alpha\beta+1}})(\alpha\beta+1)}{\gamma-1}.
\end{align}

If the condition in (\ref{eq: cond}) is satisfied, a suitable threshold for $\mathrm{Th}_1$ is given by
\begin{align}
\label{eq: Th1}
\mathrm{Th}_1=\frac{m_L+m_2}{2}=\frac{\gamma+2\alpha\beta+1}{2(\alpha\beta+1)}m_L.
\end{align}

For the separability of three high-level Gaussian distributions, a more relaxed condition should be held since the mean of the high-level Gaussian distributions, $m_H$, is far separated from the mean of the three middle Gaussian distributions. This condition is stated as
\begin{align}
\label{eq: cond1}
m_3+3\sigma_3\ll m_H-3\sqrt{0.5a\alpha\beta R_{L_0}}.
\end{align}

The condition of (\ref{eq: cond1}) is satisfied again by properly selecting the parameter $m_L$ of bias. The simplified condition of separability is given by
\begin{align}
m_L\gg \frac{3\Big(\sqrt{\frac{a\alpha\beta R_{L_0}}{\alpha+\beta}}+\sqrt{0.5a\alpha\beta R_{L_0}}\Big)}{\gamma-c_3}=k_2\sigma_{R_{L_0}},
\end{align}
where we have
\begin{align}
k_2\triangleq\frac{3\Big(\sqrt{\frac{\alpha\beta}{\alpha+\beta}}+\sqrt{0.5\alpha\beta}\Big)(\alpha+\beta)}{\alpha(\gamma-1)}.
\end{align}

In this regard, the suitable threshold of $\mathrm{Th}_2$ is defined as
\begin{align}
\label{eq: Th2}
\mathrm{Th}_2=\frac{m_H+m_3}{2}=\frac{\alpha+2\gamma\beta+\alpha\gamma}{2(\alpha+\beta)}m_L.
\end{align}

Hence, satisfying the condition $m_L>\mathrm{max}(l_1,k_2)\sigma_{R_{L_0}}$ in the design of FH-KLJN scheme, and by selecting the proper values for thresholds in (\ref{eq: Th1}) and (\ref{eq: Th2}), we can identify the 6 different Gaussian distributions from the three middle Gaussian distributions. In fact, within each chip duration, we gather enough noise samples to estimate the mean, and this mean is then compared with the two thresholds to detect the six Gaussian distributions, which, as stated earlier, are equivalent to the eight cases of 16 cases mentioned in the previous section. These are the cases that we should discard in the RH-KLJN algorithm. In this regard, after detecting the other 8 cases which are equivalent to 3 middle Gaussian distributions, we should resort to a detector to detect the first and third Gaussian distribution of among the three middle Gaussian distributions. Then, we can detect sub-bits and main bit based on the detector and our own main and sub-bit. The detector aims to distinguish between these three Gaussian distributions, which are close to each other. In the next part, we propose two detectors.

After identifying the Gaussian distribution, which is nominated as the Gaussian detection step, we should perform a bit detection step. In fact, after detecting the cases of the two Gaussian distributions of the first and third Gaussian in the middle Gaussian distributions, the bit detection step is to flip our own bit. It means that the main bit of our partner is the flip version of our own main bit, and the sub-bit of the partner is the flip of our own sub-bit. Therefore, putting all together, the final RH-KLJN algorithm is depicted in Algorithm 1. Since the minimum distance detector neglects the effect of variances and may deteriorate its performance when the means of three Gaussian distributions are close to each other, we propose ML and MAP detectors, as well as a threshold-based detector, which are presented in the detector part of Algorithm 1 and derived in the subsequent sections.

\textbf{Remark 1}: The Data Rate Improvement Factor (DRIF) is equal to $\mathrm{DRFI}=\frac{P}{2}+1$ since we hold 4 states of 16 existing states, which is equivalent to half of sub-bits ($P$ sub-bits) and half of main bits in the proposed scheme, while the classical KLJN scheme holds only half of main bits.


\subsection{ML detector}
In this part, we derive the ML detector and two other threshold-based detectors. To derive the ML detector, we assume a Hypothesis testing problem in which each hypothesis represents the existence of a specific Gaussian distribution. So, the Hypotheses are defined as
\begin{align}
\mathrm{H}_1:v_{c,n}=m_2+n_1,\\
\mathrm{H}_2:v_{c,n}=m_1+n_2,\\
\mathrm{H}_3:v_{c,n}=m_3+n_3,
\end{align}
where $n_g$ for $1\le g\le 3$ is the noise terms with zero mean and variance $\sigma^2_g$, where $\sigma^2_g$ are defined in Table II.

To derive the ML detector based on all measurement samples in the $p$'th chip duration, we define the measurement vector of common voltage as $\vb_c=[v_{c,1},v_{c,2},...,v_{c,N}]^T$, where $N$ is the number of samples in a chip duration. Then we can write the log-likelihood of each hypothesis as
\begin{align}
L_g&=\ln p(\vb_c|H_g)=\ln\prod_{n=1}^Np(v_{c,n})=\\
&\sum_{n=1}^N \ln(\frac{1}{\sigma_g\sqrt{2\pi}})-\frac{1}{2\sigma^2_g}(v_{c,n}-m_g)^2\equiv\\
&-N\ln\sigma_g-\frac{1}{2\sigma^2_g}||\vb_c-m_g||^2.
\end{align}

Therefore, maximizing $L_g$ for $1\le g\le 3$ is equivalent to minimizing $M_g=-L_g=N\ln\sigma_g+\frac{1}{2\sigma^2_g}||\vb_c-m_g||^2$. Hence, the ML detector is defined as
\begin{align}
\label{eq: MLdet}
\mathrm{ML}:\hat{g}_{\mathrm{ML}}=\mathrm{argmin}_g\quad M_g=N\ln\sigma_g+\frac{1}{2\sigma^2_g}||\vb_c-m_g||^2.
\end{align}

As can be seen, the effect of variances is taken into account in this detector.


\subsection{Threshold-based detector}
If we assume that the means of three middle Gaussian distributions are very far apart in terms of their variances, we can suggest a threshold-based detector. In this regard, the conditions of separability of the middle Gaussian distributions are
\begin{align}
\label{eq: cond3}
m_2+3\sigma_2\ll m_1-3\sigma_1,
\end{align}
\begin{align}
\label{eq: cond4}
m_1+3\sigma_1\ll m_3-3\sigma_3.
\end{align}

With some manipulations, the separability condition of (\ref{eq: cond3}) is simplified to
\begin{align}
m_L\gg \frac{3\Big(\sqrt{\frac{a\alpha R_{L_0}}{\alpha+1}}+\sqrt{\frac{a\alpha\beta}{\alpha\beta+1}R_{L_0}}\Big)}{c_1-c_2}=k_3\sigma_{R_{L_0}},
\end{align}
where we have
\begin{align}
k_3\triangleq\frac{3\Big(\sqrt{\frac{\alpha}{\alpha+1}}+\sqrt{\frac{\alpha\beta}{\alpha\beta+1}}\Big)(\alpha+\beta)(\alpha\beta+1)}{\alpha(\gamma\beta^2-\gamma-\beta^2)}.
\end{align}

Also, the separability condition of (\ref{eq: cond4}) is simplified to
\begin{align}
m_L\gg \frac{3\Big(\sqrt{\frac{a\alpha R_{L_0}}{\alpha+1}}+\sqrt{\frac{a\alpha\beta}{\alpha+\beta}R_{L_0}}\Big)}{c_3-c_1}=k_4\sigma_{R_{L_0}},
\end{align}
where we have
\begin{align}
k_4\triangleq\frac{3\Big(\sqrt{\frac{\alpha}{\alpha+1}}+\sqrt{\frac{\alpha\beta}{\alpha+\beta}}\Big)(\alpha+\beta)(\alpha+1)}{\alpha(1-\gamma-\beta+\gamma\beta)}.
\end{align}

Putting together the separability condition of the middle Gaussian distribution with the higher and lower Gaussian distributions, the overall condition of separability is defined as
\begin{align}
\label{eq: condsep}
m_L\gg \mathrm{max}(k_1,k_2,k_3,k_4)\sigma_{R_{L_0}}.
\end{align}

If the separability condition in (\ref{eq: condsep}) is held, the simple threshold detector uses the middle of the means as the thresholds, as is depicted in Figure~4. This way, the simple thresholds are given as:
\begin{align}
\mathrm{Th}_3=\frac{m_2+m_1}{2}=\frac{2\alpha^2\beta+2\gamma+\alpha+\alpha(\gamma\beta+\gamma+\beta)}{2(\alpha\beta+1)(\alpha+1)}m_L,
\end{align}
\begin{align}
\mathrm{Th}_4=\frac{m_1+m_3}{2}=\frac{2\alpha^2+2\gamma\beta+\alpha+\alpha(\gamma\beta+\gamma+\beta)}{2(\alpha+1)(\alpha+\beta)}m_L.
\end{align}

\begin{algorithm}[h]
\caption{The Proposed secure RH-KLJN algorithm in $p$'th chip duration}
\textbf{Input:}   \textbf{Common voltage samples} $v_{c_{p,n}}$; $1\le n\le N$;\\
\textbf{Parameters} $R_{H_0}$; $R_{H_1}$; $R_{L_0}$; $R_{L_1}$; $m_H$; $m_L$; $N$. \newline
\textbf{Output:} $b_{A,p}$, $b_{B,p}$. \newline
\textbf{Initialize}\\
                    $m_1=c_1m_L=\frac{\alpha+\gamma}{\alpha+1}m_L$,\\
                    $m_2=c_2m_L=\frac{\gamma+\alpha\beta}{\alpha\beta+1}m_L$,\\
                    $m_3=c_3m_L=\frac{\alpha+\gamma\beta}{\alpha+\beta}m_L$,\\
                    $\mathrm{Th}_1=\frac{m_L+m_2}{2}=\frac{\gamma+2\alpha\beta+1}{2(\alpha\beta+1)}m_L$,\\
                    $\mathrm{Th}_2=\frac{m_H+m_3}{2}=\frac{\alpha+2\gamma\beta+\alpha\gamma}{2(\alpha+\beta)}m_L$,
\label{Algorithm_1}
\begin{itemize}
\item Step 1: calculate the mean of chip duration \\$\hat{m}=\frac{1}{N}\sum_{n=1}^Nv_{c,n}$.\\
              If $\hat{m}<\mathrm{Th}_1$ or $\hat{m}>\mathrm{Th}_2$ then discard to share bits and go to step End.\\
              else
\item Step 2: \textbf{Gaussian Detector step}: \\
                \textbf{ML Detector}:\\
                $\hat{g}=\mathrm{argmin}_g\quad N\ln\sigma_g+\frac{1}{2\sigma^2_g}||\vb_c-m_g||^2$.\\
                \textbf{Threshold-based Detector}:\\
                \textbf{Simple Threshold}:\\
                $\mathrm{Th}_3=\frac{m_2+m_1}{2}=\frac{2\alpha^2\beta+2\gamma+\alpha+\alpha(\gamma\beta+\gamma+\beta)}{2(\alpha\beta+1)(\alpha+1)}m_L$\\
                $\mathrm{Th}_4=\frac{m_1+m_3}{2}=\frac{2\alpha^2+2\gamma\beta+\alpha+\alpha(\gamma\beta+\gamma+\beta)}{2(\alpha+1)(\alpha+\beta)}m_L$\\
                \textbf{Optimum Threshold}:\\
                $\mathrm{Th}_{3,opt}=\frac{-A+\sqrt{B^2-4AC}}{2A}$\\
                $\mathrm{Th}_{4,opt}=\frac{-\bar{A}+\sqrt{\bar{B}^2-4\bar{A}\bar{C}}}{2\bar{A}}$\\
                If $\hat{m}>\mathrm{Th}_4$ Then $\hat{g}=3$\\
                else if $\hat{m}>\mathrm{Th}_3$ Then $\hat{g}=1$\\
                else $\hat{g}=2$.
\item Step 3: \textbf{Bit Detector Step}:\\
                If $\hat{g}=1$ Then discard to share bits and go to step End.
                else\\
                The main bit is the flip of our own main bit $b_{A or B}=\bar{b}_{B or A}$.\\
                The main sub-bit is the flip of our sub-bit $b_{A or B,p}=\bar{b}_{B or A,p}$.
\item Step End.
\end{itemize}
\end{algorithm}
\vspace{-2ex}

If the three Gaussian distributions are not well separated, the prior probabilities of each Gaussian and its variance will be important in deriving the optimum threshold based on the minimum error probability (the error of detection between these three Gaussian distributions). Inspiring from the Binary Pulse Amplitude Modulation (B-PAM) in classical communication and decision theory, we find the optimum threshold. To obtain the threshold $\mathrm{Th}_3$, we define the error probability between the left Gaussian distribution and the middle Gaussian distribution, which is nominated as $p_{e,1}$. This is equal to
\begin{align}
p_{e,1}&=\frac{1}{4}p\{\hat{m}>\mathrm{Th}_3\|H_2\}+\frac{1}{2}p\{\hat{m}<\mathrm{Th}_3|H_1\}\\
&=\frac{1}{4}\mathrm{Q}\Big(\frac{\mathrm{Th}_3-m_2}{\sigma_2}\Big)+\frac{1}{2}\Big[1-\mathrm{Q}\Big(\frac{\mathrm{Th}_3-m_1}{\sigma_1}\Big)\Big].
\end{align}

To minimize $p_{e,1}$ with respect to $\mathrm{Th}_3$, we enforce the partial derivative equal to zero. So, we have
\begin{align}
\label{eq: Th3der}
\frac{\partial p_{e,1}}{\partial \mathrm{Th}_3}=\frac{1}{4\sigma_2}\mathrm{Q}^{'}\Big(\frac{\mathrm{Th}_3-m_2}{\sigma_2}\Big)-\frac{1}{2\sigma_1}\mathrm{Q}^{'}\Big(\frac{\mathrm{Th}_3-m_1}{\sigma_1}\Big)=0,
\end{align}
where $\mathrm{Q}(.)$ is the Q-function and we have $\mathrm{Q}^{'}(x)=\frac{-1}{\sqrt{2\pi}}\exp(\frac{x^2}{2})$. Replacing the derivative in (\ref{eq: Th3der}) and taking the logarithm of both sides, the optimum threshold is obtained from solving the quadratic problem of $Ay^2+By+C=0$, where $y=\mathrm{Th}_3$ and we have
\begin{align}
A&=1-\frac{\sigma^2_1}{\sigma^2_2},\\
B&=-2m_1+2\frac{\sigma^2_1}{\sigma^2_2}m_2,\\
C&=m^2_1-\frac{\sigma^2_1}{\sigma^2_2}m^2_2-2\sigma^2_1\ln(\frac{2\sigma_2}{\sigma_1}).
\end{align}

The optimum threshold for $\mathrm{Th}_3$ is equal to the positive root of the quadratic equation, which is
\begin{align}
\mathrm{Th}_{3,opt}=\frac{-A+\sqrt{B^2-4AC}}{2A}.
\end{align}

Note that we have $\sigma_1 < \sigma_2$, and hence $A > 0$. Similar derivations lead to the optimum threshold of $\mathrm{Th}_4$, which is equal to
\begin{align}
\mathrm{Th}_{4,opt}=\frac{-\bar{A}+\sqrt{\bar{B}^2-4\bar{A}\bar{C}}}{2\bar{A}},
\end{align}
where we have
\begin{align}
\bar{A}&=1-\frac{\sigma^2_1}{\sigma^2_3},\\
\bar{B}&=-2m_1+2\frac{\sigma^2_1}{\sigma^2_3}m_3,\\
\bar{C}&=m^2_1-\frac{\sigma^2_1}{\sigma^2_3}m^2_3-2\sigma^2_1\ln(\frac{2\sigma_3}{\sigma_1}).
\end{align}

Again, we have $\sigma_1<\sigma_3$ and hence we have $\bar{A}>0$.


\section{Analysis}
\label{sec: ana}

\subsection{Secrecy Capacity (SC) for RH-KLJN}
The fundamental metric for PLS (see, for example, \cite{Mitev23}) is the Secrecy Capacity (SC), which represents the maximum rate at which a message can be reliably transmitted while keeping it secret from an eavesdropper (Eve). For a wiretap channel model, SC is defined as:

\begin{align}
C_s = \max \{ I(X; Y) - I(X; Z), 0 \},
\end{align}
where \( I(X; Y) \) is the mutual information between Alice's transmitted signal \( X \) and Bob's received signal \( Y \), and \( I(X; Z) \) is the mutual information between \( X \) and Eve's intercepted signal \( Z \).

In the context of RH-KLJN, the "signal" \( X \) is the sequence of resistor pairs selected by Alice and Bob during each sub-bit interval, and Bob's observation \( Y \) is the common voltage \( v_c \) measured over each chip duration \( T_c \). Eve's observation \( Z \) is also \( v_c \), but without knowledge of the local sub-bits.

(a) Mutual Information for Bob (\( I(X; Y) \)):

Bob knows his own sub-bit \( s_B \) and main bit \( b_B \). For a given resistor combination \( X = (R_A, R_B) \), the common voltage \( v_c \) follows a Gaussian distribution with mean \( m_c \) and variance \( \sigma_c^2 \) (as derived in the tables 1-4). Thus, the mutual information for Bob is:
\begin{align}
I(X; Y) = H(Y) - H(Y|X),
\end{align}
where \( H(Y) \) is the entropy of \( v_c \), and \( H(Y|X) \) is the conditional entropy. Since \( v_c \) is Gaussian, its entropy is \( H(Y) = \frac{1}{2} \log(2\pi e \sigma_Y^2) \), and \( H(Y|X) = \frac{1}{2} \log(2\pi e \sigma_c^2) \). However, because Bob uses his local knowledge to resolve ambiguity, the effective \( I(X; Y) \) is high. For practical purposes, in the ideal case:

\begin{align}
I(X; Y) \approx \log_2(M),
\end{align}
where \( M \) is the number of distinguishable resistor combinations (e.g., \( M = 3 \) for the mixed states).

(b) Mutual Information for Eve (\( I(X; Z) \)):

Eve does not know the local sub-bits. She observes \( v_c \) and must distinguish between the Gaussian distributions. However, for the mixed states (01 and 10), the distributions are identical in variance and mean (in the symmetric case), so Eve cannot distinguish between them. Thus, we have

\begin{align}
I(X; Z) = 0.
\end{align}
for the mixed states. For the all-low (00) and all-high (11) states, Eve can easily identify the bits, so those cases are discarded. Therefore, the effective \( I(X; Z) \) is zero for the secure bits.

(c) Secrecy Capacity for RH-KLJN:

Since \( I(X; Z) = 0 \) for the secure bits (mixed states), the secrecy capacity simplifies to:

\begin{align}
C_s = I(X; Y) = \log_2(M).
\end{align}

For \( M = 3 \) distinguishable mixed states, \( C_s = Ln 3 \) bits per channel use. However, since only half the bits (mixed states) are used, the effective secrecy rate is:

\begin{align}
R_s = \frac{P_{\text{mixed}}}{T_b} \cdot C_s,
\end{align}

where \( P_{\text{mixed}} = 0.5 \) is the probability of mixed states, and \( T_b \) is the bit duration.

\subsection{Secrecy Outage Probability (SOP)}

The Secrecy Outage Probability (SOP) is the probability that the instantaneous secrecy capacity \( C_s \) falls below a target secrecy rate \( R_t \):

\begin{align}
\text{SOP} = \mathbb{P}(C_s < R_t).
\end{align}

In RH-KLJN, secrecy outage can occur due to:\\
- Non-ideal Gaussian separation: If the means \( m_c \) of the mixed states are not sufficiently separated, Eve might distinguish them.\\
- Imperfect parameter selection: If resistors or biases are chosen poorly, distributions may overlap.\\
- Eve's advanced capabilities: Eve might use ML detectors or exploit non-idealities.\\

Derivation of SOP:

Let \( \Delta m = \min_{i \neq j} |m_i - m_j| \) be the minimum distance between the means of the Gaussian distributions. Let \( \sigma_{\max}^2 \) be the maximum variance. The probability of Eve correctly distinguishing two distributions is related to the Q-function as:

\begin{align}
P_e^{\text{Eve}} = Q\left( \frac{\Delta m}{2 \sigma_{\max}} \right).
\end{align}

If \( P_e^{\text{Eve}} < 0.5 \), Eve cannot distinguish better than random guessing. The SOP can be expressed as:

\begin{align}
\text{SOP} = \mathbb{P}\left( \frac{\Delta m}{2 \sigma_{\max}} < \gamma \right),
\end{align}
where \( \gamma \) is a threshold related to \( R_t \). For a target \( R_t \), we require:

\begin{align}
C_s \geq R_t \implies \frac{\Delta m}{2 \sigma_{\max}} \geq \gamma_t.
\end{align}

\begin{table}[!b]
\caption{Summary of PLS equations}
\centering
 \begin{tabular}{p{30mm}||p{30mm}} \hline
 \Tstrut Matric & Equation \\\hline \hline

 \Tstrut Secrecy Capacity
 &	$\!\!\!\begin{aligned} C_s = \log_2(M)& \end{aligned}$  \\ \hline

 \Tstrut Secrecy Rate
 & $\!\!\!\begin{aligned} R_s = \frac{P_{\text{mixed}}}{T_b} \cdot C_s& \end{aligned}$ \\ \hline

 \Tstrut SOP
 &  $\mathbb{P}\left( \frac{\Delta m}{2 \sigma_{\max}} < \gamma_t \right)$ \\ \hline

 Effective Secrecy Rate
 &	$R_s^{\text{eff}} = (1 - \xi)(1 - \rho) \cdot \frac{1}{T_b} \cdot \log_2(M)$ \\ \hline

\end{tabular}
 \begin{tabular} {l}
\\
\end{tabular}
\label{Table_4}
\end{table}

Thus, we have:

\begin{align}
\text{SOP} = \mathbb{P}\left( \frac{\Delta m}{2 \sigma_{\max}} < \gamma_t \right).
\end{align}

In practice, \( \Delta m \) and \( \sigma_{\max} \) are functions of the resistor values and biases. For instance, we have:

\begin{align}
\Delta m = \min \{\left| m_1-m_2 \right|, \left| m_1-m_3 \right|\},
\end{align}

\begin{align}
\sigma_{\max}^2 = 4kT \Delta f \cdot \max(\sigma_2,\sigma_3).
\end{align}

\subsection{Ergodic Secrecy Capacity}

For a fading environment (if extended to wireless), the ergodic secrecy capacity is:

\begin{align}
\bar{C}_s = \mathbb{E}[ \max\{ I(X; Y) - I(X; Z), 0 \} ].
\end{align}

But in the wired RH-KLJN scheme, the channel is static, so this simplifies to the expressions above.

\subsection{Practical Secrecy Equation for RH-KLJN}

Incorporating practical aspects:\\
- Let \( \xi \) be the fraction of bits discarded due to non-ideal conditions.\\
- Let \( \rho \) be the fraction of bits Eve can correctly detect.\\

Then, the effective secrecy rate is:

\begin{align}
R_s^{\text{eff}} = (1 - \xi) \cdot \left( 1 - \rho \right) \cdot \frac{1}{T_b} \cdot \log_2(M),
\end{align}
where:\\
- \( \xi = \mathbb{P}(\text{non-ideal discard}) \)\\
- \( \rho = Q\left( \frac{\Delta m}{2 \sigma_{\max}} \right) \)\\

\subsection{Summary of PLS Equations for RH-KLJN}

Summary of the PLS equation for RH-KLJN is provided in Table VI.


\textbf{Remark 2: Summary of PLS analysis}\\
The PLS performance of the RH-KLJN scheme is determined by the distinguishability of the Gaussian distributions generated by resistor hopping. The secrecy capacity is maximized when the means of the distributions are well-separated relative to their variances. The SOP quantifies the probability that this separation is insufficient. Practical implementation requires careful selection of resistors and biases to ensure \( \Delta m / (2 \sigma_{\max}) \) is large enough to prevent eavesdropping.

\section{Simulation Results}
\label{sec: Simulation}
In this section, the performance of the proposed RH-KLJN is investigated. We use a bit duration of $T_b=1$ ms, which corresponds to a bit rate of $ r_b=1e3$ samples/s. Each bit duration is divided into $P=10$ chip durations, with a width of $T_c=0.1$ ms. The equivalent temperature of all resistors is equal to $T=300$ Kelvin, and the bandwidth is assumed to be $\Delta f=1$ MHz. The parameter $a$ is equal to $1.6560e-14$. The low resistors are selected as $R_{L_0}=50k$ Ohms and $R_{L_1}=\beta R_{L_0}=175k$ Ohms, where the parameter $\beta$ is assumed all over, if not otherwise stated, to be equal to $3.5$. The parameter $\alpha$ is set to $10$ all over, if not otherwise stated. The high resistors are equal to $R_{H_0}=500k$ Ohms and $R_{H_1}=1.75M$ Ohms. By tuning the bias parameter $m_L$ to $100\mu$ Volts, and selecting $\gamma=50$, the other bias is equal to $m_H=5m$ Volts. By assuming these parameters, the means of the three Gaussian distributions are equal to $m_1=5.4545e-4$, $m_2=2.36113e-4$, and $m_3=1.4e-3$, respectively. The standard deviations of the Gaussian distributions are equal to $\sigma_1=2.7436e-5$, $\sigma_2=2.8373e-5$, and $\sigma_3=4.6332e-5$. As we can observe, the Gaussian distributions are well-separated. We nominate this case as a good example of separability. In contrast, if we choose a smaller bias of $m_L=95\mu$ Volts, then the means are $m_1=5.1818e-4$, $m_2=2.2431e-4$, and $m_3=1.3e-3$, while the standard deviations are the same as before. In this case, the means are closer to each other, and the separability is slightly violated. We nominated this case as a moderate separability case. The number of main bits is assumed to be $K=4e7$, and hence the number of sub-bits is equal to $4e8$ bits. The performance metric of the RH-KLJN communication system is the BEP which is defined as the ratio of error detected bits of Bob by Alice in the total bit exchange bits. Five experiments are performed. The number of samples in a chip duration is assumed to be $N=20$, if not otherwise stated.

\begin{figure}[tb]
\begin{center}
\includegraphics[width=9cm]{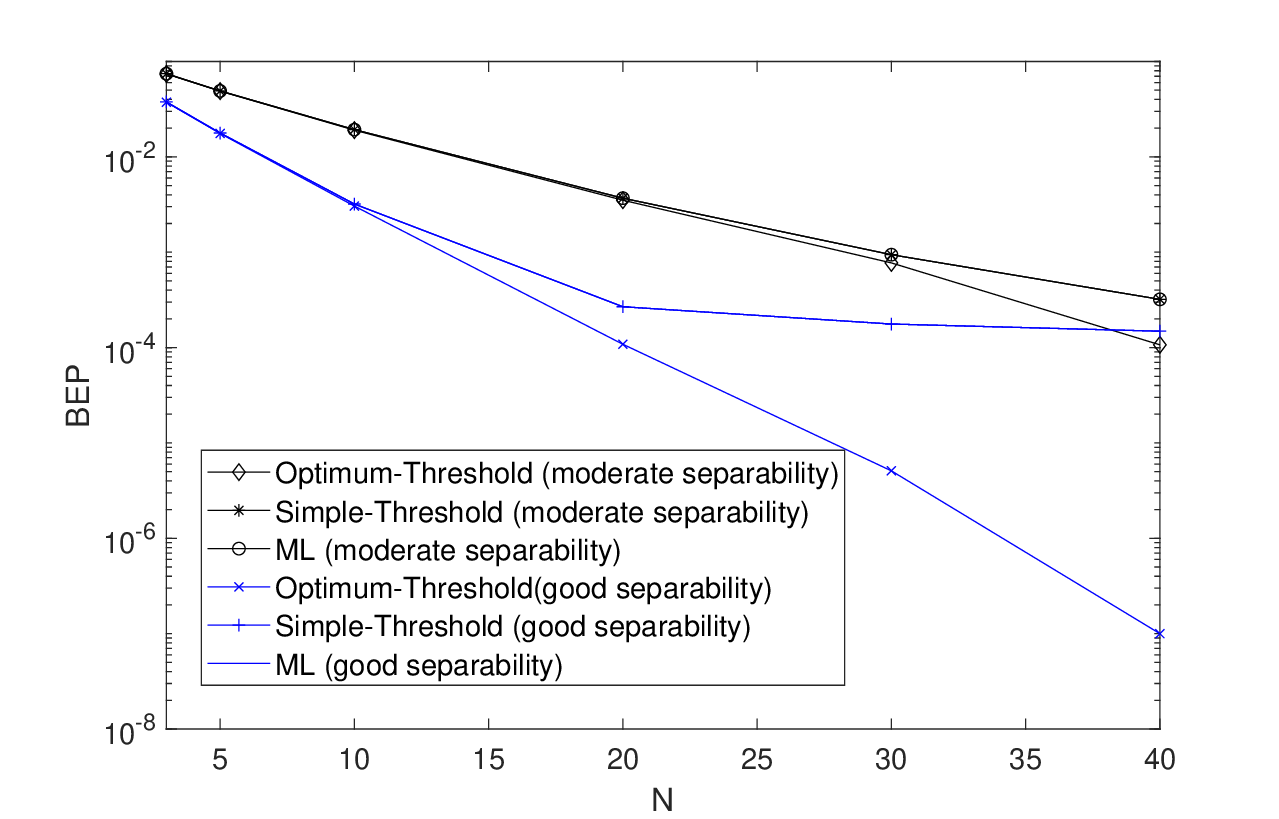}
\end{center}
\vspace{-0.5 cm}
\caption{BEP versus number of chip samples $N$.}
\label{figN}
\end{figure}

In the first experiment, the effect of $N$ on the performance of the proposed RH-KLJN is investigated using three detectors: an ML detector, a simple threshold-based detector, and an optimum threshold-based detector, in both good separability cases and moderate separability cases. The number $N$ is varied from 3 to 40. The BEP versus $N$ is depicted in Fig.~\ref{figN}. It can be seen that the simple threshold-based detector and the ML detector have the same BEP, while the optimum threshold-based detector has a lower BEP, especially in larger values of $N$. Also, the good separability case has a lower BEP than the moderate separability case. Moreover, as evident from the increasing sampling rate by increasing $N$, the BEP decreases. In all experiments, as we will see, the simple detector has the same result as the ML detector.

\begin{figure}[tb]
\begin{center}
\includegraphics[width=9cm]{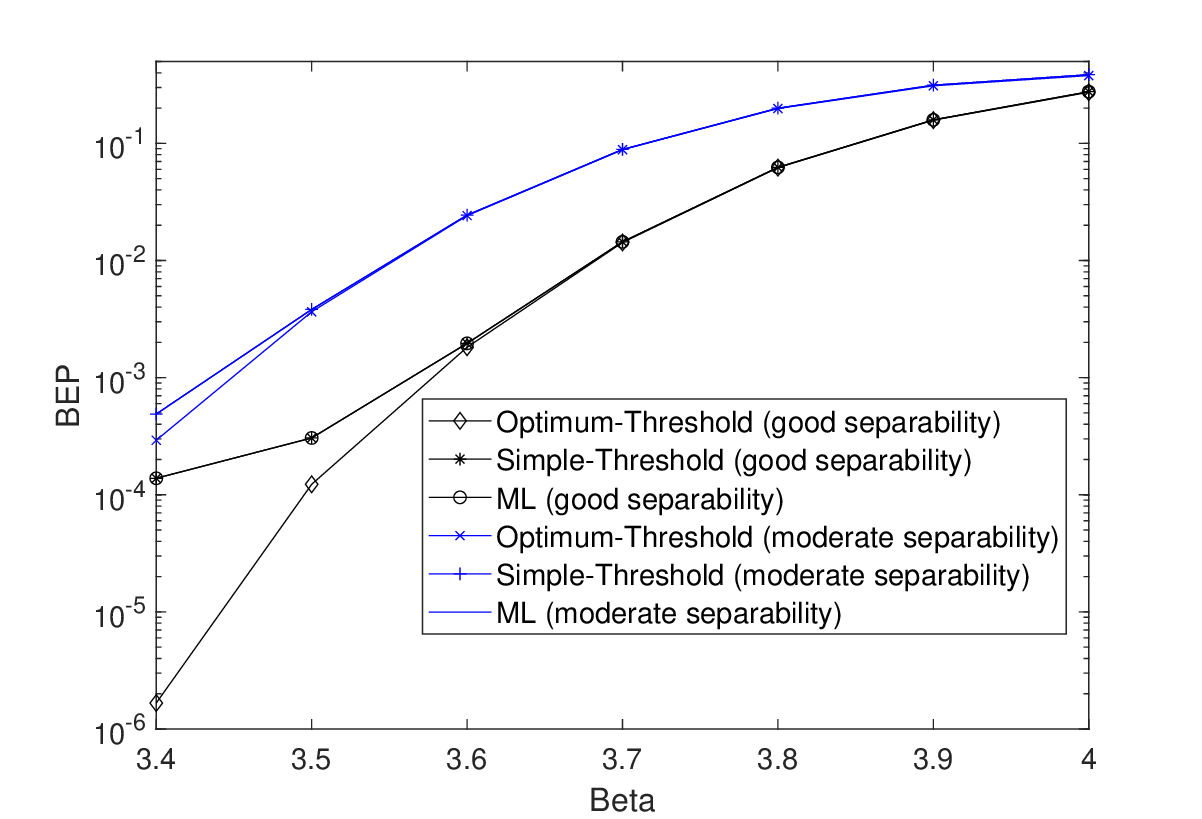}
\end{center}
\vspace{-0.5 cm}
\caption{BEP versus $\beta$.}
\label{figbeta}
\end{figure}

\begin{figure}[tb]
\begin{center}
\includegraphics[width=9cm]{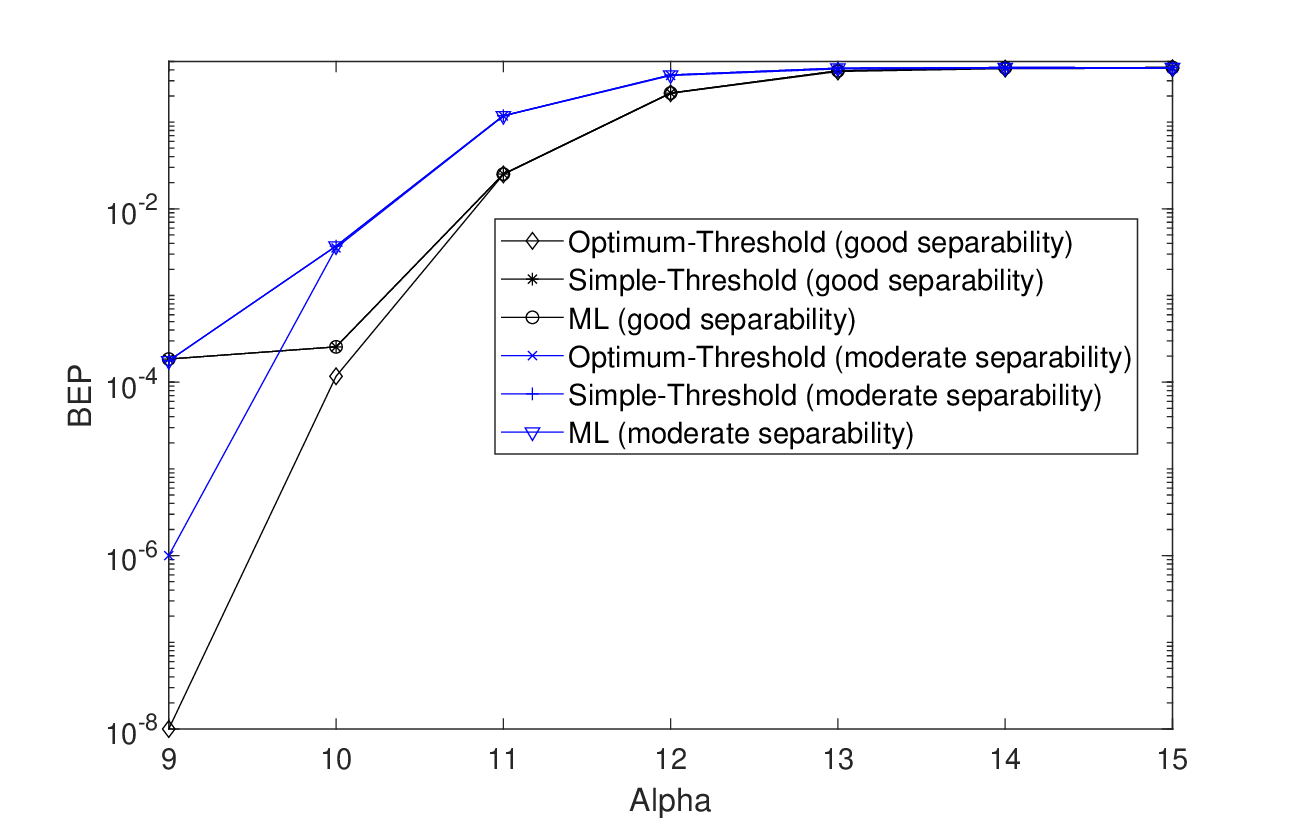}
\end{center}
\vspace{-0.5 cm}
\caption{BEP versus $\alpha$.}
\label{figalpha}
\end{figure}
\begin{figure}[tb]
\begin{center}
\includegraphics[width=9cm]{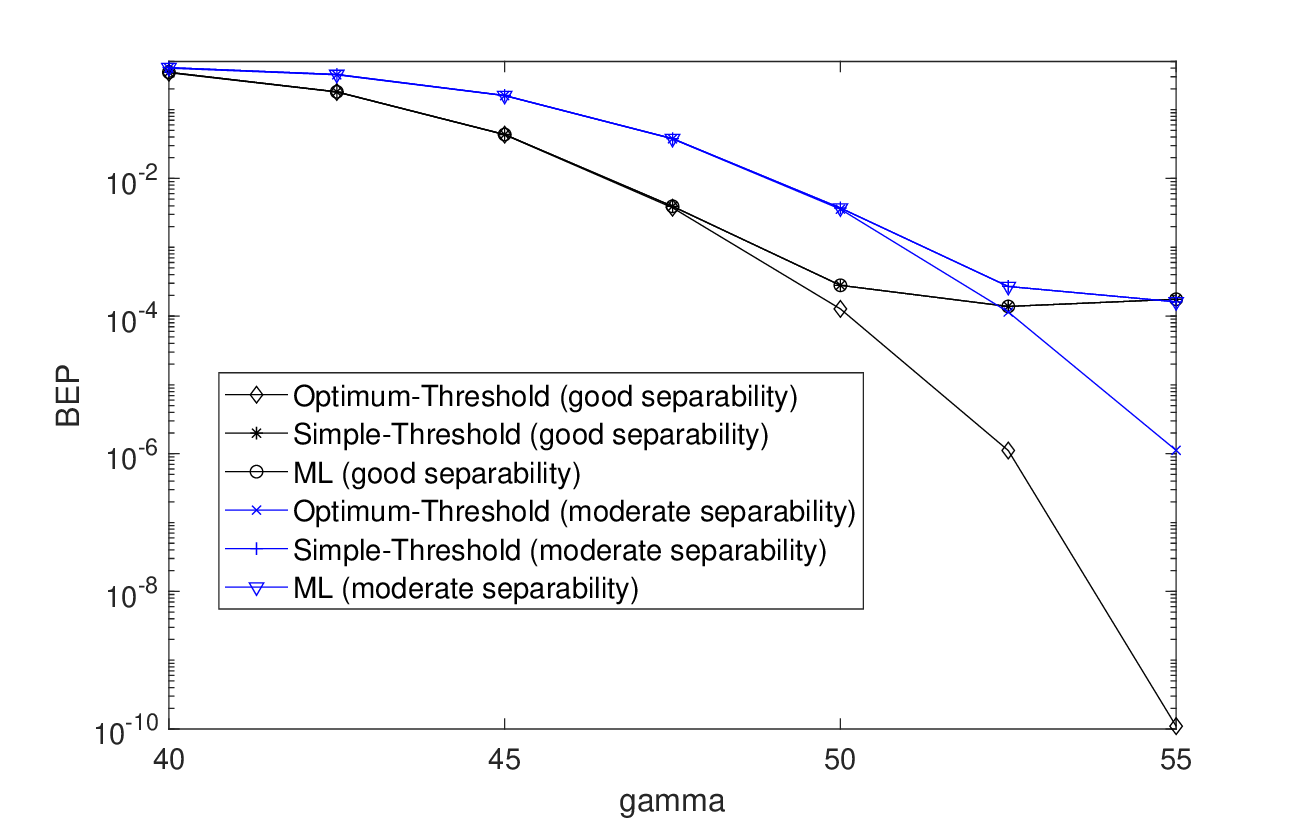}
\end{center}
\vspace{-0.5 cm}
\caption{BEP versus $\gamma$.}
\label{figgamma}
\end{figure}

In the second experiment, the effect of parameter $\beta$ is investigated. The parameter $\beta$ is varied between 3.4 and 4. This range is completely sweep the performance of the system from low BEP to high BEP and is obtained by a trial and error. The BEP versus $\beta$ is shown in Fig.~\ref{figbeta}. It can be observed that increasing the value of $\beta$ results in an increase in the BEP. In other words, the value of $\beta$ should be far smaller than the value of $\alpha=10$. Additionally, the good separability case has a lower BEP than the moderate separability case, and the optimum threshold-based detector yields a better result than both the simple threshold-based detector and the ML detector.

In the third experiment, the effect of parameter $\alpha$ is discussed. The parameter $\alpha$ is varied between 9 and 15. This range is completely sweep the performance of the system from low BEP to high BEP and is obtained by trial and error. The BEP versus $\alpha$ is demonstrated in Fig.~\ref{figalpha}. It can be deduced that by selecting $\beta=3.5$, the value of $\alpha$ should not be much greater than $\beta$. The results are the same as those in the previous experiment.

\begin{figure}[tb]
\begin{center}
\includegraphics[width=9cm]{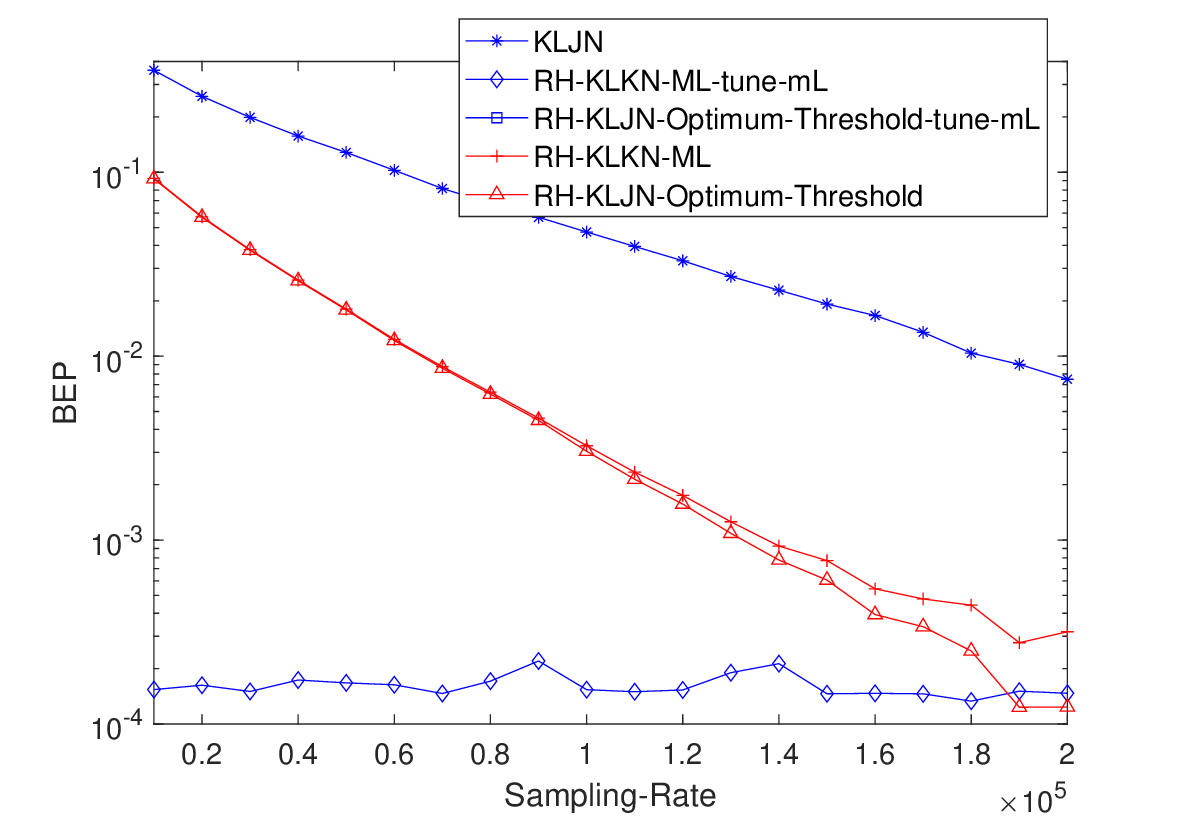}
\end{center}
\vspace{-0.5 cm}
\caption{BEP versus sampling rate for KLJN and RH-KLJN.}
\label{figKLJN}
\end{figure}

In the fourth experiment, the effect of parameter $\gamma$ is investigated. The parameter $\gamma$ is selected between 40 and 55. This range is completely sweep the performance of the system from low BEP to high BEP and is found by a trial and error. The BEP versus $\gamma$ is depicted in Fig.~\ref{figgamma}. It can be observed that the BEP decreases as $\gamma$ increases. Thus, the value of $\gamma$ should be kept large enough. The same results also hold in this experiment.

In the fifth experiment, we compare the proposed RH-KLNJN scheme with the classical KLJN. To fairly compare the two schemes, we use the same sampling rate for both. In other words, we consider the same number of samples in each bit duration in both schemes. We defined the sampling rate as $\mathrm{SR}_{KLJN}=\frac{N_{KLJN}}{T_b}$ for the KLJN approach and $\mathrm{SR}_{RH-KJLN}=\frac{N_{RH-KLJN}}{PT_b}$ for the RH-KLJN approach. Moreover, to see the effect of fine tuning the value of $m_L=10\mathrm{max}(k_1,k_2,k_3,k_4)\sigma_{R_{L_0}}\gg \mathrm{max}(k_1,k_2,k_3,k_4)\sigma_{R_{L_0}}$ in the performance of the proposed approach, we also depict this case in addition to the default case of $m_L=1e-4$ with good separability. BEP versus sampling rate for the fine-tuned case, and the good separability case is shown in Fig.~\ref{figKLJN}. It shows that the proposed RH-KLJN scheme has at least one decade and somewhere two decades lower BEP than KLJN in a fair comparison. In addition to better performance of the proposed RH-KLJN in terms of lower BEP, the proposed scheme has a data-rate improvement factor of $\mathrm{DRIF}=\frac{10}{2}+1=6$ in comparison to the classical KLJN scheme, as mentioned in Remark 1 of the paper. It verifies the superiority of the proposed scheme both in improving the secure data rate and enhancing the quality of the communication system in terms of BEP. Indeed, it is achieved by increasing the complexity of the communication system and utilizing low-voltage biases, resulting in a power-efficient scheme compared to a zero-power communication system. Moreover, Fig.~\ref {figKLJN} demonstrates the benefit of deliberately fine-tuning the value of $m_L$ to achieve a lower BEP in comparison to the ad-hoc good separable case. Additionally, the figure shows that the two cases, fine-tuned and good separability, achieve the same performance when the sampling rate is sufficiently large. Therefore, the fine-tuned case yields a lower BEP than the good separable case, with an even lower sampling rate. It shows the importance of separability conditions, which are derived in (\ref{eq: condsep}).



\section{Conclusion and future work}
\label{sec: con}
This paper generalizes the concept of frequency hopping in classical communication theory to resistor hopping in a secure KLJN noise communication framework. The proposed RH-scheme enables a significantly higher data rate between Alice and Bob by utilizing a large number of chip durations within a bit duration. It also leverages the BEP in comparison to the classical KLJN scheme, assuming both systems have the same sampling rate. The simulation results confirm these results. Moreover, the paper presents an ML detector and a minimum error probability-based threshold detector with the optimum threshold calculated in a closed form. Additionally, the separability condition is discussed, which enables the selection of the bias voltage of $m_L$ in a closed form. This fine-tuned selection of parameter $m_L$ exhibits a significant reduction in BEP compared to other ad-hoc selections of this bias. Therefore, the separability of the Gaussian distributions is possible and verified through simulations and by properly selecting the system's parameters. As analyzed in the PLS equations, separability is a requisite for security, which can be achieved. Finding the exact calculation of BEP and generalizing the hopping to more than two resistors (which would be very difficult to consider in general) is left for future work.

\end{document}